\documentclass[lettersize,journal]{IEEEtran}
\usepackage{amsmath,amsfonts}
\usepackage{physics,amsmath}
\usepackage{algorithm}
\usepackage{algpseudocode}
\usepackage{multirow}
\usepackage{array}
\usepackage{bm}
\usepackage{placeins}
\usepackage {wrapfig}
\usepackage{xcolor}
\usepackage{tcolorbox}
\usepackage{amsfonts}

\usepackage{amsmath,amssymb}

\usepackage{subcaption}
\usepackage{mathtools}
\usepackage{xcolor}
\usepackage{float} 
\usepackage{textcomp}
\usepackage{stfloats}
\usepackage{amsmath}   

\usepackage{lipsum}  
\usepackage{verbatim}
\usepackage{graphicx}
\usepackage{pdfpages}
\usepackage{tikz}
\usepackage{lettrine}
\usetikzlibrary{3d}
\usetikzlibrary{arrows.meta} 

\usepackage{tikz}
\usetikzlibrary{shapes, arrows.meta, positioning, calc, decorations.pathreplacing, patterns, fit, backgrounds}
\usepackage{amsmath, amssymb}
\usepackage[colorlinks=true, linkcolor=blue, urlcolor=blue, citecolor=blue, anchorcolor=blue]{hyperref}

\begin{document}

\title{Performance Limits of Hardware-Constrained THz Inter-Satellite MIMO-ISAC Systems}
\author{Haofan Dong,~\IEEEmembership{Student Member,~IEEE}, 
        and Ozgur B. Akan,~\IEEEmembership{Fellow,~IEEE}
\thanks{The authors are with Internet of Everything Group, Department of Engineering, University of Cambridge, CB3 0FA Cambridge, UK.}
\thanks{Ozgur B. Akan is also with the Center for neXt-generation Communications
(CXC), Department of Electrical and Electronics Engineering, Koç University, 34450 Istanbul, Turkey (email:oba21@cam.ac.uk)}}
	   



\maketitle

\begin{abstract}
Terahertz inter-satellite links (THz-ISL) offer unprecedented bandwidth for future space networks but face fundamental constraints from onboard power and thermal budgets. This paper establishes theoretical performance limits for MIMO Integrated Sensing and Communication (ISAC) systems under per-element constant-envelope (CE) transmission constraints. We demonstrate that hardware distortions---specifically power amplifier nonlinearity, ADC quantization, and oscillator phase noise---impose a capacity ceiling that cannot be overcome by increasing transmit power. A unified link budget framework integrates wideband beam squint, aperture pointing errors, and colored noise sources through a spectral consistency principle that ensures residual phase noise is counted exactly once across communication and sensing analyses. The sensing bounds are derived via the Whittle-Fisher Information Matrix under a Constant Acceleration kinematic model with jerk noise, yielding closed-form scaling laws: residual phase noise variance scales as $\alpha^{-1}$ while dynamic state-estimation error (DSE) variance scales as $\alpha^{-5}$ with pilot overhead $\alpha$. Numerical results show divergent MIMO scaling: sensing precision improves with array size ($\mathrm{RMSE} \propto 1/\sqrt{N_t N_r}$), while the critical SNR exhibits scale invariance regarding array size, implying that the distortion-limited transition point stabilizes regardless of the array scale.
The steep $\alpha^{-5}$ DSE scaling creates an operationally infeasible region at $\alpha < \alpha^* \approx 0.16$, where $\alpha^* = (C_{\mathrm{DSE}}/C_{\mathrm{PN}})^{1/4}$---a constraint-driven threshold under the adopted baseline for LEO operation. These findings provide design guidelines for hardware-efficient THz-ISL constellations.
\end{abstract}

\begin{IEEEkeywords}
Terahertz communications, inter-satellite links, ISAC, MIMO, hardware impairments, Cramér-Rao bound, phase noise, beam squint.
\end{IEEEkeywords}

\definecolor{ieee_blue}{RGB}{0, 80, 155}
\definecolor{ieee_red}{RGB}{180, 30, 30} 
\definecolor{ieee_gray}{RGB}{128, 128, 128}

\section{Introduction}\label{sec:introduction}
The deployment of Low Earth Orbit (LEO) mega-constellations represents a pivotal architecture for next-generation non-terrestrial networks (NTN), aiming to provide seamless global coverage and ultra-high-speed connectivity~\cite{dureppagari2023ntn, dureppagari2025leo}. To satisfy the multi-Tbps capacity requirements of Inter-Satellite Links (ISLs) while ensuring autonomous operation, the integration of sensing and communications (ISAC) has emerged as a critical enabler. ISAC capabilities allow satellites to utilize shared spectrum and hardware for simultaneous high-speed data transmission and high-precision tasks such as relative positioning for formation flying and space debris detection~\cite{dong2025debrisense}. The terahertz (THz) band (0.1--10 THz) is particularly advantageous for this application, offering vast bandwidths for capacity and sub-millimeter wavelengths for ultra-high-resolution sensing~\cite{jiang2024terahertz, dong2025martian}. However, the severe path loss associated with THz propagation necessitates the use of massive Multiple-Input Multiple-Output (MIMO) arrays to generate sufficient beamforming gain ($g_{\rm ar}$) to close the link budget.

\begin{figure*}[t]
  \centering
  \includegraphics[width=0.9\linewidth]{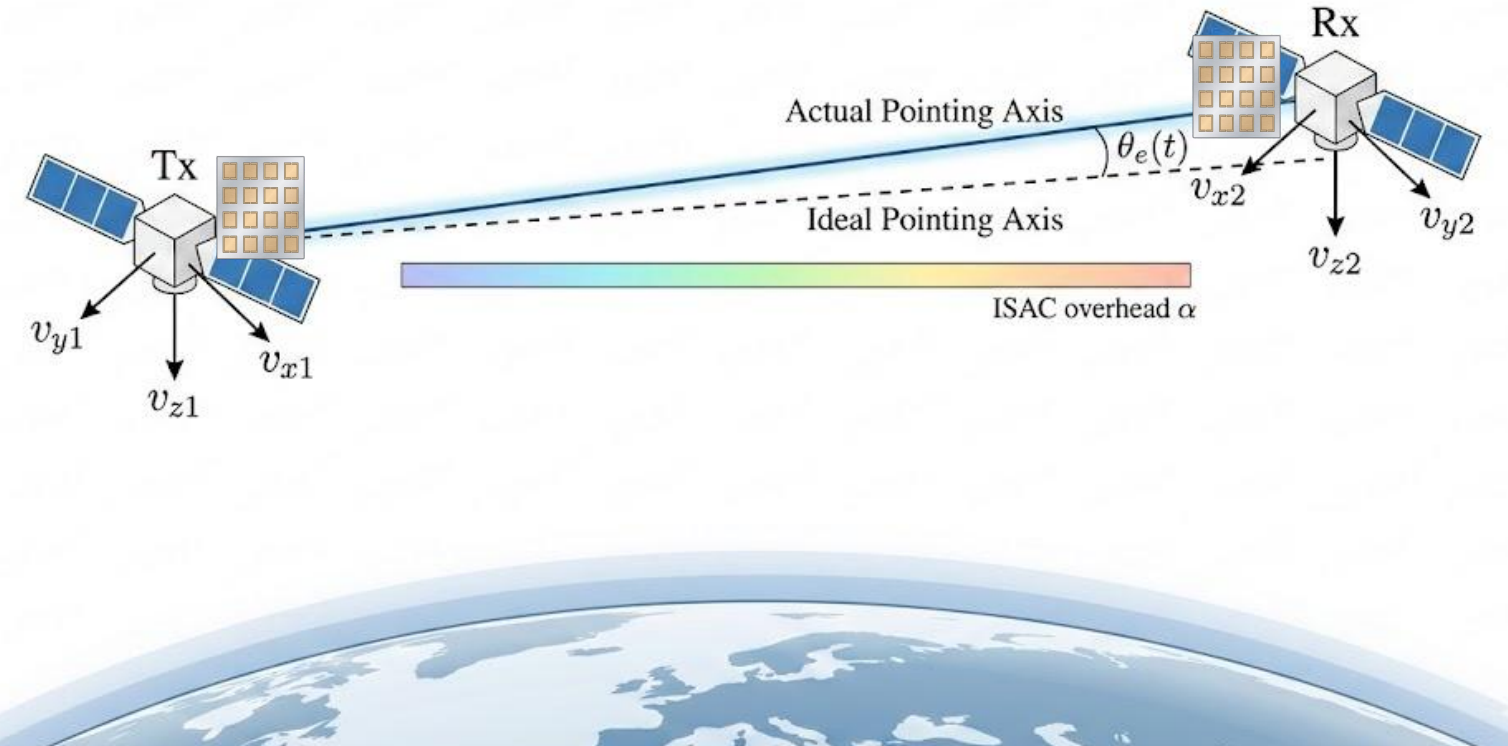}%
  \caption{Geometry of the LEO--LEO THz ISL with pointing error and joint ISAC operation. The Tx/Rx body frames $(v_{x1},v_{y1},v_{z1})$ and $(v_{x2},v_{y2},v_{z2})$ are shown, together with the ideal and actual pointing axes, the time-varying error $\theta_e(t)$ at the receiver, and the ISAC overhead~$\alpha$.}
  \label{fig:system_geometry}
\end{figure*}

While the theoretical potential of THz MIMO ISAC is significant, its implementation on LEO platforms is fundamentally constrained by the strict Size, Weight, and Power (SWaP) limitations of the space environment. Unlike terrestrial base stations, satellites operate under rigorous Direct Current (DC) power budgets and thermal management constraints restricted to radiative heat dissipation~\cite{chen2025review, solyman2021potential}. These constraints impose a ``Constant Envelope (CE) Firewall,'' mandating High Power Amplifiers (HPAs) to operate near saturation to maximize DC-to-RF conversion efficiency~\cite{jiang2013effect}. Consequently, conventional linear MIMO precoding schemes, such as Singular Value Decomposition (SVD) or Zero-Forcing (ZF), become practically infeasible due to their inherently high Peak-to-Average Power Ratio (PAPR)~\cite{liu2009error}. Under the stringent Per-PA Power Constraint (PAPC)~\cite{song2010quantization}, high PAPR signals necessitate large Input Back-Off (IBO), resulting in either unacceptable power efficiency degradation~\cite{xie2023dynamic, burla2013integrated} or severe signal distortion when IBO is reduced~\cite{bae2017use}. Here, distortion is quantified by a composite hardware quality factor, $\Gamma_{\rm eff}$~\cite{dong2025fundamental,dong2025fundamental2}, which encompasses PA nonlinearity, quantization errors, and phase noise. As illustrated in Fig.~\ref{fig:system_geometry}, we consider a THz inter-satellite link between two LEO platforms, where a narrow THz beam is steered along the nominal line-of-sight (LoS) axis. Platform micro-vibrations introduce small but non-negligible pointing errors between the ideal and actual beam axes, while untracked lethal debris fragments in the beam corridor generate weak forward-scattered fields that can be exploited for joint communication and sensing.

This operational reality necessitates a structural adaptation: rather than pursuing spatial multiplexing, the array gain is leveraged to combat the elevated noise floor from hardware impairments ($\Gamma_{\rm eff}$), effectively employing an equivalent single-stream architecture~\cite{tsinos2020constant}. Establishing the fundamental limits of this hardware-constrained CE-MIMO architecture requires resolving three analytical challenges typically treated in isolation: (i) coupling between wideband beam squint $\eta_{\rm bsq}(f)$ and colored noise $N(f)$, which invalidates conventional FIM formulations; (ii) the dual role of common phase noise---multiplicative coherence loss for communication versus additive noise for sensing~\cite{aminu2019beamforming,khanzadi2015capacity}; and (iii) consistent $\Gamma_{\rm eff}$ normalization when distortion scales with frequency-dependent array gain~\cite{dong2025fundamental}.

To address these issues, this paper develops a unified dual-scale framework integrating communication-side coherence loss (Bussgang model) with sensing-side residual noise (Whittle-FIM). The contributions are:

\begin{itemize}
\item \textbf{Constraint-Driven CE-MIMO Framework:} A physical layer model for THz-ISL under per-element CE constraint, resolving the $\Gamma_{\rm eff}$ normalization and enforcing spectral consistency for phase noise single-counting.

\item \textbf{Hardware-Limited Capacity Analysis:} A closed-form Jensen bound ($C_{\rm J}^{\rm MIMO}$) via Bussgang decomposition, quantifying the saturation ceiling $C_{\rm sat}$ and the critical SNR transition to distortion-limited operation.

\item \textbf{Sensing Bounds under Colored Noise:} Dual FIM interfaces (Whittle-FIM and exact time-domain) for wideband parameter estimation, with Misspecified CRB for DSE characterization.

\item \textbf{Operationally Infeasible Region:} The $\alpha^{-5}$ DSE variance scaling from jerk-noise accumulation creates a minimum pilot overhead $\alpha^* = (C_{\mathrm{DSE}}/C_{\mathrm{PN}})^{1/4}$, below which system operation becomes impractical.

\end{itemize}

\section{System Model and Hardware Constraints}
\label{sec:system_model}

\begin{figure*}[!t]
    \centering
    \includegraphics[width=\textwidth]{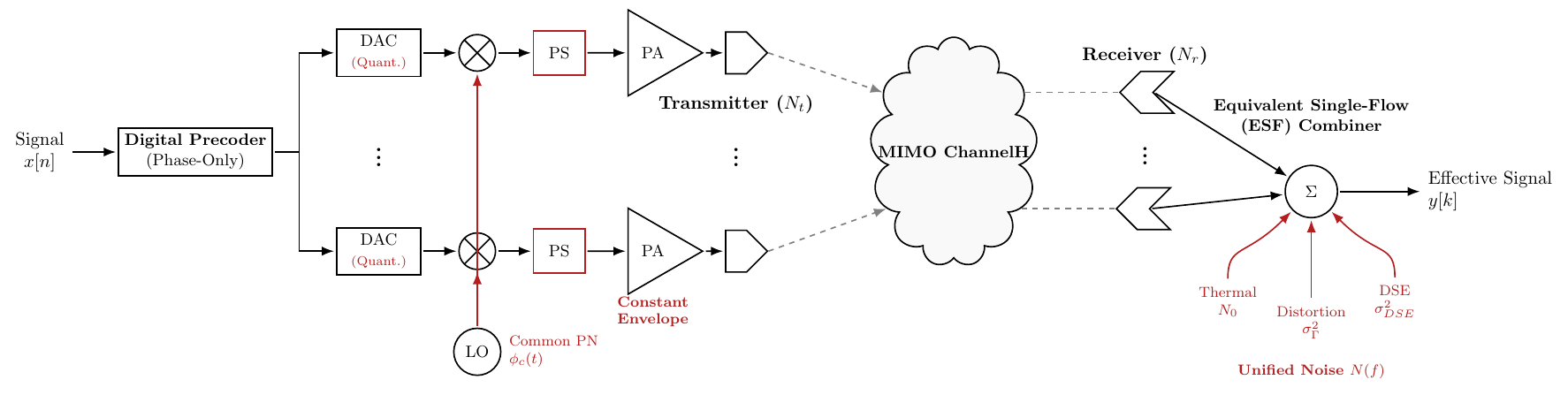}
    \caption{Hardware-constrained THz-ISL CE-MIMO-ISAC transceiver. A phase-only precoder under per-PA constant-envelope constraints drives the DAC/mixer/PS/PA chains, while a shared LO with common phase noise $\phi_c(t)$ feeds all RF paths. At the receiver, coherent combining yields an equivalent single flow (ESF) with unified coloured noise spectrum $N(f)$ at the output $y[k]$.}
    \label{fig:ce_mimo_esf}
\end{figure*}

This section establishes a unified physical layer model for THz-ISL MIMO-ISAC systems operating under realistic hardware constraints. The model comprises three interconnected components: (i) a multiplicative signal gain framework capturing geometric and hardware-induced coherence losses (Section~\ref{sub:signal_model}), (ii) an additive noise and distortion model with explicit treatment of colored noise sources (Section~\ref{sub:noise_model}), and (iii) a resource allocation model governing noise variance scaling with pilot overhead (Section~\ref{sub:resource_model}). A \emph{spectral consistency principle} ensures that residual phase noise is counted exactly once across communication and sensing analyses. The resulting hardware-constrained CE-MIMO-ISAC transceiver architecture is summarized in Fig.~\ref{fig:ce_mimo_esf}, which highlights the phase-only digital precoder, per-element RF chains, shared local oscillator, and the equivalent single-flow (ESF) combiner that aggregates all distortions into a unified noise spectrum.

\subsection{Conventions and Array Geometry}
\label{sub:conventions}

The baseband signal is observed over duration $T$ with $N$ samples at interval $T_s = T/N$. The bandwidth $B$ satisfies $B \leq 1/T_s$, and the DFT employs $N$ frequency bins with spacing $\Delta f = B/N$. The Parseval normalization $\sum_{n=0}^{N-1} |x[n]|^2 = N^{-1} \sum_{k=0}^{N-1} |X[k]|^2$ ensures energy consistency for FIM derivations.

A uniform linear array (ULA) with $N_t$ transmit and $N_r$ receive elements is assumed. The apertures are $L_{\mathrm{ap,tx}} = N_t d$ and $L_{\mathrm{ap,rx}} = N_r d$, with $L_{\mathrm{ap}}$ denoting the active aperture when unambiguous. The far-field condition requires $R \gg 2 L_{\mathrm{ap}}^2 / \lambda_c$. The beam steering angle $\theta_0$ is measured from broadside, with the $n$-th element phase shift $\phi_n = -2\pi n d \sin\theta_0 / \lambda_c$ referenced to $f_c$.

\subsection{Unified Signal and Multiplicative Gain Model}
\label{sub:signal_model}

The effective array gain $G_{\mathrm{sig}}$ determines both communication capacity and sensing CRB. The ideal coherent gain $G_{\mathrm{ideal}} = N_t N_r$ is degraded by a cascade of statistically independent multiplicative loss factors. The baseline propagation includes free-space path loss $(4\pi R/\lambda_c)^2$ and atmospheric absorption $G_{\mathrm{path}}(f)$, treated as known constants.

\textit{Equivalent Single-Flow Model with LEO Dynamics:}
Given the per-element constant-envelope (CE) constraint imposed by THz PA design, the MIMO channel reduces to an equivalent single-flow (ESF) architecture where gain is realized exclusively through phase-coherent array combining. To capture the high-mobility LEO environment, the discrete-frequency baseband signal at the $k$-th bin is:
\begin{equation}
    Y[k] = \sqrt{P_{\mathrm{rx}}} \cdot S[k; \boldsymbol{\theta}] \cdot e^{j 2\pi f_D k T_s} \cdot e^{j\phi_{c,\mathrm{res}}[k]} + W[k],
    \label{eq:esf_model}
\end{equation}
where $P_{\mathrm{rx}} = P_{\mathrm{tx}} G_{\mathrm{sig,avg}}$ is the effective received power, $f_D$ is the Doppler shift induced by relative satellite motion (up to $\pm 7$~MHz at 140~GHz for LEO-LEO links with $\pm 15$~km/s relative velocity), $S[k; \boldsymbol{\theta}]$ is the normalized transmitted spectrum with parameter vector $\boldsymbol{\theta} = [\tau, f_D]^T$, $\phi_{c,\mathrm{res}}[k]$ is the residual common phase noise after tracking, and $W[k]$ is the aggregate additive noise. The carrier tracking loop is designed with $B_L = 10$~MHz to accommodate this Doppler range while maintaining acceptable residual phase noise.

\textit{Beam Squint Loss:}
The first and often dominant loss factor stems from the incompatibility between phase-only (PO) beamforming and wideband signaling. In a PO architecture, each element applies a fixed phase $\phi_n(f_c)$ optimized for coherent combining at $f_c$. At any off-center frequency $f \neq f_c$, the phase mismatch $\Delta\phi_n(f) = 2\pi n d \sin\theta_0 (f - f_c)/c$ accumulates linearly across the aperture, causing the beam to deviate from the intended direction---a phenomenon known as beam squint.

For symmetric Tx/Rx PO architectures in the ISL scenario, the total link gain degradation is the product of transmit and receive losses, yielding the fourth-power form:
\begin{equation}
    \eta_{\mathrm{bsq}}(f) = \operatorname{sinc}^4 \left( \pi \frac{L_{\mathrm{ap}}}{\lambda_c} \sin\theta_0 \frac{f - f_c}{f_c} \right).
    \label{eq:eta_bsq_k}
\end{equation}
This squared effect relative to single-ended radar systems imposes a stringent constraint on the permissible bandwidth-aperture product. The in-band average $\bar{\eta}_{\mathrm{bsq}} = B^{-1} \int_{f_c - B/2}^{f_c + B/2} \eta_{\mathrm{bsq}}(f) \, df$ admits a Taylor expansion. Using $\operatorname{sinc}^4(x) \approx 1 - 2\pi^2 x^2/9$ for $|x| \ll 1$:
\begin{align}
    \bar{\eta}_{\mathrm{bsq}} &\approx 1 - \frac{2\pi^2}{9} \left(\frac{\pi L_{\mathrm{ap}} \sin\theta_0}{\lambda_c f_c}\right)^2 \cdot \frac{1}{B}\int_{-B/2}^{B/2} f^2 \, df \notag \\
    &= 1 - \frac{\pi^2}{18} \left( \frac{L_{\mathrm{ap}}}{\lambda_c} \sin\theta_0 \right)^2 \left( \frac{B}{f_c} \right)^2, \quad \kappa \lesssim 0.5,
    \label{eq:eta_bsq_approx}
\end{align}
where $\kappa = (L_{\mathrm{ap}}/\lambda_c)(B/f_c)|\sin\theta_0|$. This formula reveals that coherence loss scales quadratically with the space-bandwidth product, establishing a fundamental aperture-bandwidth limit. For the reference configuration ($L_{\mathrm{ap}}/\lambda_c = 50$, $B/f_c = 0.067$, $\theta_0 = 30°$), $\kappa = 1.675$ exceeds the validity region; numerical integration of~\eqref{eq:eta_bsq_k} is employed. The threshold $\kappa \approx 1$ ($\sim$1~dB loss) marks where TTD networks (2--4~dB insertion loss) become cost-effective.

\textit{Static Hardware Losses:}
Beyond beam squint, static hardware impairments introduce additional multiplicative losses. For $b_\phi$-bit phase shifters with quantization step $\Delta\phi = 2\pi/2^{b_\phi}$:
\begin{equation}
    \rho_Q = \operatorname{sinc}^2 \left( \frac{\pi}{2^{b_\phi}} \right).
    \label{eq:rho_Q}
\end{equation}
For typical implementations with $b_\phi \geq 4$, this loss remains below 0.2~dB. However, in low-cost designs where $b_\phi = 2$--$3$, the penalty can exceed 1~dB, necessitating trade-offs between power consumption (which scales exponentially with $b_\phi$) and link budget margin.

Platform micro-vibrations induced by mechanical disturbances (reaction wheels, solar panel deployment) and thermal cycling introduce random angular perturbations $\Delta\theta \sim \mathcal{N}(0, \sigma_\theta^2)$. Following the Ruze exponential degradation model~\cite{ruze1966antenna}:
\begin{equation}
    \rho_{\mathrm{APE}} = \exp\left(-\frac{\pi^2}{3} \left( \frac{L_{\mathrm{ap}} \cos\theta_0}{\lambda_c} \right)^2 \sigma_\theta^2\right).
    \label{eq:rho_APE}
\end{equation}
For a normalized aperture of $L_{\mathrm{ap}}/\lambda_c = 1000$ with $\sigma_\theta = 0.5$~mrad, this yields $\rho_{\mathrm{APE}} \approx 0.57$ ($\sim$2.4~dB loss), underscoring the necessity of active platform stabilization to maintain $\sigma_\theta \lesssim 0.1$~mrad for THz-ISL applications.

\textit{Remark (Distinction from Ruze Equation):} The coefficient $\pi^2/3$ in~\eqref{eq:rho_APE} differs from the $4\pi^2$ factor in the standard Ruze equation. This distinction arises because the Ruze model addresses \emph{uncorrelated} surface errors that destroy phase coherence across the aperture, whereas the present model describes \emph{correlated} platform jitter that induces a linear phase gradient (beam misalignment). The coefficient $\pi^2/3$ is derived from a second-order Taylor expansion of the array $\mathrm{sinc}^2$ radiation pattern under Gaussian pointing statistics.

For amplitude errors $A_n = 1 + \epsilon_n$ with $\epsilon_n \sim \mathcal{N}(0, \sigma_a^2)$, the coherent power loss follows the Ruze degradation model:
\begin{equation}
    \rho_A \approx e^{-\sigma_a^2}, \quad \sigma_a \ll 1.
    \label{eq:rho_A}
\end{equation}
For typical $\sigma_a = 0.1$ (10\% RMS error), the loss is $\sim$0.04~dB; for $\sigma_a = 0.2$, it increases to $\sim$0.17~dB.

\textit{Differential Phase Noise Loss:}
In shared-LO architectures, the phase noise at each element decomposes as $\phi_n(t) = \phi_c(t) + \delta\phi_n(t)$. The common component $\phi_c(t)$, originating from the central reference oscillator, induces synchronized constellation rotation that can be partially tracked by carrier recovery. The differential component $\delta\phi_n(t)$, arising from distributed PLLs and per-channel jitter, destroys phase coherence among array elements. The coherent \emph{power} factor (not amplitude) for Gaussian $\delta\phi_n$ with variance $\sigma_{\mathrm{rel}}^2$ is:
\begin{equation}
    \rho_{\mathrm{PN}} = \left| \frac{1}{N} \sum_{n=1}^{N} \mathbb{E}\left[ e^{j\delta\phi_n} \right] \right|^2 = \left| e^{-\sigma_{\mathrm{rel}}^2/2} \right|^2 = e^{-\sigma_{\mathrm{rel}}^2},
    \label{eq:rho_PN}
\end{equation}
with $\sigma_{\mathrm{rel}}^2 = \sigma_{\mathrm{rel,tx}}^2 + \sigma_{\mathrm{rel,rx}}^2 + (2\pi f_c \sigma_{t,\mathrm{diff}})^2$, where the last term converts differential timing jitter $\sigma_{t,\mathrm{diff}}$ to phase variance at carrier frequency $f_c$. Critically, only $\delta\phi_n$ contributes to the multiplicative loss; the tracking residual of $\phi_c(t)$ appears as additive noise (Section~\ref{sub:noise_model}).

\textit{Unified Gain:}
Aggregating all independent loss factors, the total effective gain is:
\begin{equation}
    G_{\mathrm{sig, avg}} = G_{\mathrm{ideal}} \cdot \bar{\eta}_{\mathrm{bsq}} \cdot \rho_Q \cdot \rho_{\mathrm{APE}} \cdot \rho_A \cdot \rho_{\mathrm{PN}}.
    \label{eq:G_sig_avg}
\end{equation}
For sensing FIM computation requiring frequency-resolved characteristics, the signal amplitude spectrum is:
\begin{equation}
    s(f, \boldsymbol{\theta}) \propto \sqrt{G_{\mathrm{ideal}} \cdot \rho_Q \cdot \rho_{\mathrm{APE}} \cdot \rho_A \cdot \rho_{\mathrm{PN}} \cdot \eta_{\mathrm{bsq}}(f)}.
    \label{eq:s_f_theta}
\end{equation}

\subsection{Unified Noise and Distortion Model}
\label{sub:noise_model}

The additive noise PSD $N(f)$ comprises white and colored components. A \emph{spectral consistency principle} ensures the residual common phase noise $\phi_{c,\mathrm{res}}(t)$ is counted exactly once per analysis domain.

\textit{Hardware Distortion:}
The total distortion power ratio sums independent contributions~\cite{bae2017use}:
\begin{equation}
    \Gamma_{\mathrm{eff, total}} = \Gamma_{\mathrm{PA}} + \Gamma_{\mathrm{ADC}} + \Gamma_{\mathrm{IQ}} + \Gamma_{\mathrm{LO}}.
    \label{eq:Gamma_eff_total}
\end{equation}
Each component captures a distinct hardware impairment with characteristic magnitude:
\begin{itemize}
    \item $\Gamma_{\mathrm{PA}}$: PA nonlinearity (EVM floor), typically $-30$ to $-20$~dB for class-AB amplifiers at mmWave frequencies under CE operation.
    \item $\Gamma_{\mathrm{ADC}} = 10^{-(6.02 b_{\mathrm{ADC}} + 1.76)/10}$: ADC quantization noise using the standard ideal-$b_{\mathrm{ADC}}$ SNR formula 
$ \mathrm{SNR}_{\mathrm{ADC}} = 6.02 b_{\mathrm{ADC}} + 1.76~\mathrm{dB}$~\cite{bennett1948spectra};
for $b_{\mathrm{ADC}} = 6$~bits (ENOB), this gives $\Gamma_{\mathrm{ADC}} \approx -38$~dB.
    \item $\Gamma_{\mathrm{IQ}} = 10^{-\mathrm{IRR}_{\mathrm{dB}}/10}$: I/Q imbalance where $\mathrm{IRR}_{\mathrm{dB}}$ is the image rejection ratio (positive, typically 30--40~dB for direct-conversion receivers).
    \item $\Gamma_{\mathrm{LO}} = (2\pi f_{\mathrm{eff}} \sigma_t)^2$: Clock jitter noise where $f_{\mathrm{eff}}$ is the effective signal frequency; for $\sigma_t = 50$~fs and $f_{\mathrm{eff}} = 5$~GHz, $\Gamma_{\mathrm{LO}} \approx -56$~dB.
\end{itemize}
The equivalent noise PSD is $\sigma^2_{\Gamma} = P_{\mathrm{tx}} G_{\mathrm{sig, avg}} \Gamma_{\mathrm{eff, total}} / B$. This normalization establishes a causal dependency: the signal gain calculation~\eqref{eq:G_sig_avg} precedes noise PSD computation.

\textit{Colored Noise Sources:}
Two frequency-dependent noise sources distinguish the model from classical AWGN. Residual sideband modulation $S_{\mathrm{RSM}}(f)$ arises from the dual-use ISAC waveform that embeds communication data through continuous phase modulation (CPM) on the radar chirp carrier. While the receiver's matched filter suppresses CPM spectral sidelobes, imperfect suppression allows residual energy to leak into the sensing bandwidth. The PSD $S_{\mathrm{RSM}}(f)$ exhibits peaks at symbol rate harmonics and nulls at frequencies determined by the modulation index. In the time domain, this yields a non-diagonal Toeplitz noise covariance matrix $\boldsymbol{\Sigma}_{\mathrm{RSM}}$, necessitating the Whittle-FIM frequency-domain framework developed in Section~\ref{subsec:sensing_bounds}.

Residual common phase noise after carrier tracking with loop bandwidth $B_L$ (defined as the 3-dB frequency of the closed-loop transfer function) is:
\begin{equation}
    S_{\phi,c,\mathrm{res}}(f) = \frac{f^2}{B_L^2 + f^2} \cdot S_{\phi,c,\mathrm{tot}}(f),
    \label{eq:S_phi_c_res}
\end{equation}
where $S_{\phi,c,\mathrm{tot}}(f) = S_{\phi,c,\mathrm{tx}}(f) + S_{\phi,c,\mathrm{rx}}(f)$. The high-pass characteristic reflects that first-order PLL tracking suppresses low-frequency drift but passes high-frequency noise. For LEO-LEO links with relative velocities up to $\pm 15$~km/s, the Doppler shift reaches $\pm 7$~MHz at 140~GHz. To track this dynamic range, a loop bandwidth of $B_L = 10$~MHz is adopted, providing sufficient margin while limiting residual phase noise power to acceptable levels.

\textit{Spectral Consistency Principle:}
A critical modeling convention is required to avoid counting the same physical effect twice. The residual common phase noise $\phi_{c,\mathrm{res}}(t)$ impacts system performance through a single mechanism---phase uncertainty---but this mechanism manifests differently in communication and sensing analyses.

For \emph{communication capacity analysis}, the receiver employs carrier phase tracking to align the received constellation. Imperfect tracking introduces random phase rotation on constellation points, reducing the effective SNR for coherent detection. This effect is naturally captured by a multiplicative coherence loss $\rho_{\phi,c,\mathrm{res}} = e^{-\sigma^2_{\phi,c,\mathrm{res}}}$, where $\sigma^2_{\phi,c,\mathrm{res}} = \int_{-B/2}^{B/2} S_{\phi,c,\mathrm{res}}(f) \, df$ (Parseval's theorem). Under this convention, the additive noise PSD excludes phase noise:
\begin{equation}
    N_{\mathrm{comm}}(f) = N_0 + \sigma^2_{\Gamma} + \sigma^2_{\mathrm{DSE}}(\alpha) + S_{\mathrm{RSM}}(f).
    \label{eq:N_comm}
\end{equation}

For \emph{sensing FIM analysis}, the Cramér-Rao bound quantifies estimation performance on the raw baseband signal \emph{before} any carrier tracking is applied. From the estimator's perspective, the unknown phase process $\phi_{c,\mathrm{res}}(t)$ corrupts the parameter-bearing signal structure and must be treated as additive distortion. Furthermore, the FIM formulation requires a linear signal-in-noise model; the multiplicative coherence factor would violate this linearity. Under this convention, the noise PSD includes the residual phase noise spectrum:
\begin{equation}
    N_{\mathrm{sense}}(f) = N_{\mathrm{comm}}(f) + S_{\phi,c,\mathrm{res}}(f).
    \label{eq:N_sense}
\end{equation}

The mathematical equivalence of these two conventions follows from Parseval's theorem: the phase noise contributes $P_{\mathrm{sig}}(1 - e^{-\sigma^2_{\phi,c,\mathrm{res}}}) \approx P_{\mathrm{sig}} \sigma^2_{\phi,c,\mathrm{res}}$ as signal power loss in the communication convention, which equals the integrated additive phase noise power $\int S_{\phi,c,\mathrm{res}}(f) df = \sigma^2_{\phi,c,\mathrm{res}}$ in the sensing convention under small-phase-error approximation ($\sigma^2_{\phi,c,\mathrm{res}} \ll 1$). This duality reflects the physical distinction between \emph{coherent detection} (phase-sensitive) and \emph{parameter estimation} (extracting timing/Doppler information).

\subsection{Resource Allocation and Noise Scaling}
\label{sub:resource_model}

The sensing resource fraction $\alpha \in (0, 1]$ governs ISAC trade-offs under a constant-energy baseline ($E_{\mathrm{total}} = P_{\mathrm{tx}} T$ fixed). Two scaling laws characterize the $\alpha$-dependent noise~\cite{delos2018system}:

\textit{Residual Phase Noise Scaling:} Based on the Cramér-Rao lower bound for phase estimation~\cite{kay1993fundamentals}, the tracking quality improves with pilot energy, yielding:
\begin{equation}
    \sigma^2_{\phi,c,\mathrm{res}}(\alpha) \approx \frac{C_{\mathrm{PN}}}{\alpha},
    \label{eq:scaling_pn}
\end{equation}
where $C_{\mathrm{PN}} = \Gamma_{\mathrm{loop}} N_0 / (2 E_{\mathrm{total}})$ aggregates loop implementation loss $\Gamma_{\mathrm{loop}}$, noise floor, and total energy budget.

\textit{DSE Mismatch Scaling:} To capture LEO orbital dynamics, target state evolution is modeled using a Constant Acceleration (CA) kinematic model~\cite{bar2001estimation}. Integrating white jerk noise three times over the tracking interval $\Delta t \propto 1/\alpha$ yields a position error variance $\sigma^2_r(\Delta t) \propto \Delta t^5$; mapping to a two-way carrier phase mismatch gives:
\begin{equation}
    \sigma^2_{\phi,\mathrm{DSE}}(\alpha) \approx \frac{C_{\mathrm{DSE}}}{\alpha^5},
    \label{eq:scaling_dse}
\end{equation}
where $\sigma^2_{\phi,\mathrm{DSE}}$ denotes the phase-mismatch variance ($\mathrm{rad}^2$); the detailed derivation and normalization convention appear in Section~\ref{subsec:resource_allocation}. This steep $\alpha^{-5}$ dependency implies that DSE dominates in the low-overhead regime, rendering operation at $\alpha \ll 0.1$ impractical under the adopted baseline.

\textit{Reference SNR:}
The reference SNR is defined as $\mathrm{SNR}_0 = P_{\mathrm{tx}}/(N_0 B)$; colored additive terms ($\Gamma_{\mathrm{eff}}$, DSE, RSM, and residual PN for sensing) enter explicitly via the unified PSD models in the SINR~\eqref{eq:sinr_f} and FIM interfaces.

The unified model established in this section---comprising the multiplicative signal gain framework~\eqref{eq:G_sig_avg}, the additive noise model~\eqref{eq:N_comm}--\eqref{eq:N_sense}, and the scaling laws~\eqref{eq:scaling_pn}--\eqref{eq:scaling_dse}---provides the link budget foundation for the performance limits derived in Section~\ref{sec:performance_limits}.

\section{Performance Limits of Hardware-Constrained ISAC Systems}
\label{sec:performance_limits}

This section derives theoretical performance bounds for communication and sensing under realistic hardware impairments. The analysis implements the spectral consistency principle established in Section~\ref{sub:noise_model} to avoid double-counting common distortion sources.

\subsection{Communication Capacity Bounds}
\label{subsec:comm_capacity}

\textit{CE-MIMO Single-Stream Equivalent Model:}
The LoS-dominant THz ISL scenario exhibits low spatial rank. The rank-1 channel with normalized array response vectors $\mathbf{a}_t(\vartheta_t)$ and $\mathbf{a}_r(\vartheta_r)$ satisfying $\|\mathbf{a}_t\|_2 = \|\mathbf{a}_r\|_2 = 1$ yields the factorized channel matrix:
\begin{equation}
    \mathbf{H} \approx \sqrt{G_{\mathrm{path}}} \, \mathbf{a}_r(\vartheta_r) \mathbf{a}_t(\vartheta_t)^H.
    \label{eq:channel_los}
\end{equation}
Under the constant-envelope (CE) per-antenna constraint imposed by THz PA limitations, the transmit vector is constrained to $\mathbf{x} = \sqrt{P_{\mathrm{tx}}/N_t} \, [e^{j\theta_1}, \dots, e^{j\theta_{N_t}}]^T$ with $|x_n|^2 = P_{\mathrm{tx}}/N_t$. Phase-only precoding aligns transmit phases with the channel by setting $\theta_n = \arg([\mathbf{a}_t]_n)$, yielding:
\begin{equation}
    \mathbf{a}_t^H \mathbf{x} = \sqrt{\frac{P_{\mathrm{tx}}}{N_t}} \sum_{n=1}^{N_t} |[\mathbf{a}_t]_n| \xrightarrow{N_t \to \infty} \sqrt{P_{\mathrm{tx}}},
    \label{eq:ce_array_gain}
\end{equation}
demonstrating that the CE loss vanishes asymptotically for large arrays.

\textit{Phase Noise Linearization via Bussgang Decomposition:}
The residual phase error $\phi_{\mathrm{res}} \sim \mathcal{N}(0, \sigma^2_{\phi,c,\mathrm{res}})$ arises from imperfect pilot-based phase tracking. For Gaussian phase noise, the characteristic function of $e^{j\phi}$ yields:
\begin{equation}
    \mathbb{E}\{e^{j\phi_{\mathrm{res}}}\} = \int_{-\infty}^{\infty} e^{j\phi} \cdot \frac{1}{\sqrt{2\pi\sigma^2}} e^{-\phi^2/(2\sigma^2)} d\phi = e^{-\sigma^2_{\phi,c,\mathrm{res}}/2}.
    \label{eq:characteristic_function}
\end{equation}
Using the standard Bussgang decomposition~\cite{bussgang1952crosscorrelation} to linearize the multiplicative phase noise $e^{j\phi_{\mathrm{res}}}$, the signal separates into coherent and uncorrelated distortion terms:
\begin{align}
    e^{j\phi_{\mathrm{res}}} s_k &= \underbrace{e^{-\sigma^2_{\phi,c,\mathrm{res}}/2}}_{\text{amplitude factor}} s_k + u_k, \label{eq:bussgang_decomp} \\
    \mathbb{E}\{u_k s_k^*\} &= 0, \quad \mathbb{E}|u_k|^2 = (1 - e^{-\sigma^2_{\phi,c,\mathrm{res}}}) \mathbb{E}|s_k|^2.
\end{align}
The coherent \emph{power} factor is $|e^{-\sigma^2/2}|^2 = e^{-\sigma^2_{\phi,c,\mathrm{res}}}$, which appears in the SINR numerator.

\textit{Frequency-Dependent SINR:}
Incorporating the multiplicative gain from Section~\ref{sub:signal_model} and following the communication convention where $\phi_{c,\mathrm{res}}$ enters as multiplicative coherence loss, the SINR at frequency $f$ is:
\begin{equation}
    \mathrm{SINR}(f) = \frac{\mathrm{SNR}_0 \, G_{\mathrm{sig}}(f) \, \rho_{\mathrm{static}} \, e^{-\sigma^2_{\phi,c,\mathrm{res}}}}{1 + \mathrm{SNR}_0 \, G_{\mathrm{sig}}(f) \, \Gamma_{\mathrm{eff,total}} + \sigma^2_{\mathrm{DSE}}/N_0 + S_{\mathrm{RSM}}(f)/N_0},
    \label{eq:sinr_f}
\end{equation}
where $G_{\mathrm{sig}}(f) = G_{\mathrm{ideal}} \cdot \eta_{\mathrm{bsq}}(f)$ includes ideal array gain and beam squint loss, $\rho_{\mathrm{static}} = \rho_Q \cdot \rho_{\mathrm{APE}} \cdot \rho_A \cdot \rho_{\mathrm{PN}}$ aggregates static coherence factors, and $\mathrm{SNR}_0 = P_{\mathrm{tx}}/(N_0 B)$ is the reference SNR. The denominator separates thermal noise (unity), multiplicative distortion ($\mathrm{SNR}_0 G_{\mathrm{sig}} \Gamma_{\mathrm{eff,total}}$), and additive colored terms (DSE, RSM normalized by $N_0$).

The spectral efficiency under Gaussian-input mutual information is computed via frequency-domain integration:
\begin{equation}
    C_{\mathrm{exact}} = \frac{1}{B} \int_{f_c - B/2}^{f_c + B/2} \log_2\big(1 + \mathrm{SINR}(f)\big) \, df.
    \label{eq:capacity_exact}
\end{equation}
This expression serves as a capacity surrogate under the effective-SINR model with aggregate impairments; throughout this section, ``capacity'' refers to this Gaussian-input spectral efficiency metric.

\textit{Jensen Upper Bound:}
Since $\log_2(1+x)$ is concave, Jensen's inequality yields a tractable upper bound $C_{\mathrm{exact}} \leq C_{\mathrm{J}}$ using the band-averaged effective SINR:
\begin{equation}
    C_{\mathrm{J}} = \log_2(1 + \overline{\mathrm{SINR}}_{\mathrm{eff}}),
    \label{eq:jensen_bound}
\end{equation}
where
\begin{equation}
    \overline{\mathrm{SINR}}_{\mathrm{eff}} = \frac{\mathrm{SNR}_0 \, G_{\mathrm{sig,avg}} \, e^{-\sigma^2_{\phi,c,\mathrm{res}}}}{1 + \mathrm{SNR}_0 \, G_{\mathrm{sig,avg}} \, \Gamma_{\mathrm{eff,total}} + \sigma^2_{\mathrm{DSE}}/N_0 + \bar{S}_{\mathrm{RSM}}/N_0},
    \label{eq:sinr_avg}
\end{equation}
with $G_{\mathrm{sig,avg}} = B^{-1} \int G_{\mathrm{sig}}(f) \rho_{\mathrm{static}} \, df$. The Jensen gap is controlled by the variance of $\mathrm{SINR}(f)$. Let $g(x) = \log_2(1+x)$ with $g''(x) = -1/[(1+x)^2 \ln 2]$. A second-order Taylor expansion around $\mu = \mathbb{E}[\mathrm{SINR}]$ yields:
\begin{align}
    \mathbb{E}[g(X)] &\approx g(\mu) + \frac{1}{2}g''(\mu)\mathrm{Var}[X] = g(\mu) - \frac{\mathrm{Var}[X]}{2(1+\mu)^2 \ln 2}, \notag \\
    \Rightarrow \; \Delta_{\mathrm{J}} &= C_{\mathrm{J}} - C_{\mathrm{exact}} \approx \frac{\mathrm{Var}_f[\mathrm{SINR}(f)]}{2(1 + \mathbb{E}_f[\mathrm{SINR}(f)])^2 \ln 2}.
    \label{eq:jensen_gap}
\end{align}
Within the TTD threshold regime where $\bar{\eta}_{\mathrm{bsq}} \gtrsim -1$~dB, the beam squint variance exhibits quadratic suppression $\mathrm{Var}_f[\eta_{\mathrm{bsq}}(f)] \propto (1 - \bar{\eta}_{\mathrm{bsq}})^2$, yielding negligible $\Delta_{\mathrm{J}}$. The gap quantification is presented in Section~IV. This quadratic approximation provides analytical insight into the bandwidth-aperture scaling but is accurate only for $\kappa \lesssim 0.5$.

\textit{Saturation Capacity and Critical SNR:}
At high SNR where multiplicative distortions dominate additive terms, the effective SINR approaches:
\begin{equation}
    \overline{\mathrm{SINR}}_{\mathrm{eff}} \xrightarrow{\mathrm{SNR}_0 \to \infty} \frac{e^{-\sigma^2_{\phi,c,\mathrm{res}}}}{\Gamma_{\mathrm{eff,total}}},
\end{equation}
yielding the saturation ceiling:
\begin{equation}
    C_{\mathrm{sat}} = \log_2\left(1 + \frac{e^{-\sigma^2_{\phi,c,\mathrm{res}}}}{\Gamma_{\mathrm{eff,total}}}\right).
    \label{eq:saturation}
\end{equation}
This hardware-imposed ceiling is independent of transmit power and array size under the per-PA power constraint. For state-of-the-art hardware with $\Gamma_{\mathrm{eff,total}} \approx 6 \times 10^{-3}$ and $\sigma^2_{\phi,c,\mathrm{res}} = 0.01$, the capacity saturates at approximately 7.6~bits/s/Hz.

The critical SNR marking the transition from noise-limited to distortion-limited regime is obtained by equating noise and distortion power:
\begin{equation}
    \mathrm{SNR}_{\mathrm{crit}} = \frac{1}{G_{\mathrm{sig,avg}} \Gamma_{\mathrm{eff,total}}}.
    \label{eq:critical_snr}
\end{equation}

\textit{MIMO Scaling Physics:} In conventional MIMO with uncorrelated hardware impairments, $\Gamma_{\mathrm{eff,total}}$ remains constant with array size, and~\eqref{eq:critical_snr} would predict $\mathrm{SNR}_{\mathrm{crit}} \propto 1/(N_t N_r)$. However, under CE transmission with PA-dominated distortion, the per-element distortion beamforms coherently with the signal (see Remark in Section~\ref{subsec:comm_performance}), causing effective distortion power to scale as $N_t N_r$---identical to signal power. This ``directional distortion'' phenomenon implies $\Gamma_{\mathrm{eff,total}}^{\mathrm{(eff)}} \propto G_{\mathrm{sig,avg}}$, rendering~\eqref{eq:critical_snr} approximately constant with array size. Consequently, the distortion-limited transition point exhibits scale invariance: massive arrays improve sensing precision without forcing the communication link into saturation at lower per-element SNR.

\subsection{Sensing Performance Bounds}
\label{subsec:sensing_bounds}

\textit{Unified Noise Covariance Framework:}
Following the sensing convention, residual phase noise $\sigma^2_{\phi,c,\mathrm{res}}$ appears exclusively as additive colored noise in the total noise PSD:
\begin{equation}
    N(f) = N_0 + \sigma^2_{\Gamma} + \sigma^2_{\mathrm{DSE}} + S_{\mathrm{RSM}}(f) + S_{\phi,c,\mathrm{res}}(f),
    \label{eq:noise_psd_sense}
\end{equation}
where $\sigma^2_{\Gamma} = \Gamma_{\mathrm{eff,total}} P_{\mathrm{sig,ref}}/B$ converts multiplicative distortions to equivalent additive noise PSD, with $P_{\mathrm{sig,ref}} = P_{\mathrm{tx}} G_{\mathrm{sig,avg}} / L_{\mathrm{path}}$ denoting the reference received signal power at the combiner output. The signal model for parameter estimation with $\boldsymbol{\theta} = [\tau, f_d]^T$ is:
\begin{equation}
    S(f, \boldsymbol{\theta}) = A(f) \cdot G_{\mathrm{sig}}(f) \cdot \rho_{\mathrm{static}} \cdot s_{\mathrm{bb}}(f - f_c, \boldsymbol{\theta}),
    \label{eq:signal_sensing}
\end{equation}
where $A(f) = \sqrt{\eta_{\mathrm{bsq}}(f) \cdot \rho_{\mathrm{static}}}$ is the frequency-dependent amplitude shaping factor incorporating beam squint and static hardware losses, and the multiplicative coherence loss $e^{-\sigma^2_{\phi,c,\mathrm{res}}}$ is intentionally excluded to maintain single-counting.

\textit{Whittle-FIM Derivation:}
For colored Gaussian noise in the frequency domain, the observation at the $k$-th bin is $Y_k = S_k(\boldsymbol{\theta}) + N_k$ with $\mathbb{E}\{N_k N_{k'}^*\} = \delta_{kk'} N_k$. The Whittle-FIM provides an asymptotically correct frequency-domain formulation~\cite{kay1993fundamentals}. For delay parameter $\tau$ with signal model $S(f) = A(f) e^{-j2\pi f\tau}$, the derivative yields:
\begin{equation}
    \frac{\partial S(f)}{\partial \tau} = -j 2\pi f \, S(f).
    \label{eq:signal_derivative}
\end{equation}
The Whittle-FIM for delay estimation is:
\begin{equation}
    \mathcal{J}_{\tau\tau} = 2 \sum_k \frac{|\partial S_k / \partial \tau|^2}{N_k} \xrightarrow{\Delta f \to 0} 8\pi^2 \int_{-B/2}^{B/2} \frac{f^2 |S(f)|^2}{N(f)} \, df,
    \label{eq:whittle_fim}
\end{equation}
where the discrete sum converges to the integral as frequency spacing $\Delta f = B/N \to 0$. The $1/N(f)$ weighting adaptively suppresses high-noise frequency regions. The CRB for unbiased delay estimation is $\mathrm{CRB}_\tau = 1/\mathcal{J}_{\tau\tau}$, and the ranging RMSE follows from $r = c\tau/2$:
\begin{equation}
    \mathrm{RMSE}_r = \frac{c}{2} \sqrt{\mathrm{CRB}_\tau}.
    \label{eq:rmse_range}
\end{equation}
Here the factor $1/2$ reflects a cooperative two-way time-transfer protocol over the ISL, where both terminals transmit timestamps and the round-trip delay is obtained from two one-way active transmissions rather than passive monostatic echo; hence no additional reflection loss is incurred.

\textit{AWGN Closed-Form Baseline:}
For rectangular spectrum $|S(f)|^2 = E/B$ (constant PSD over bandwidth $B$ with total energy $E$) and white noise $N(f) = N_0$, the FIM integral evaluates as:
\begin{align}
    \mathcal{J}_{\tau\tau} &= 8\pi^2 \frac{E}{B N_0} \int_{-B/2}^{B/2} f^2 \, df = 8\pi^2 \frac{E}{B N_0} \cdot \frac{2}{3}\left(\frac{B}{2}\right)^3 \notag \\
    &= 8\pi^2 \frac{E}{B N_0} \cdot \frac{B^3}{12} = \frac{2\pi^2 E B^2}{3 N_0}. \label{eq:J_awgn}
\end{align}
The corresponding ranging RMSE is:
\begin{equation}
    \mathrm{RMSE}_r^{\mathrm{AWGN}} = \frac{c}{2} \sqrt{\frac{3 N_0}{2\pi^2 E B^2}}.
    \label{eq:awgn_baseline}
\end{equation}
This closed-form expression establishes the thermal-noise-limited baseline and reveals the fundamental $1/B$ scaling with bandwidth and $1/\sqrt{E}$ scaling with energy.

\textit{General BCRLB for Colored Noise:}
For frequency-dependent $N(f;\alpha)$, define the weighted second spectral moment:
\begin{equation}
    \mathcal{M}_2^{(-1)}(B,\alpha) \triangleq \int_{-B/2}^{B/2} \frac{f^2 W(f)}{N(f;\alpha)} \, df, \quad W(f) \triangleq \frac{|S(f)|^2}{E/B},
    \label{eq:weighted_moment}
\end{equation}
where $W(f)$ is the normalized signal power distribution. The general BCRLB expression becomes:
\begin{equation}
    \mathrm{BCRLB}(\alpha) = \frac{c^2}{32\pi^2} \cdot \frac{B}{E} \cdot \frac{1}{\mathcal{M}_2^{(-1)}(B,\alpha)}.
    \label{eq:bcrlb_general}
\end{equation}
The complete FIM incorporates prior information: $\mathbf{F}_{\mathrm{total}} = \mathbf{F}_D + \mathbf{F}_P$, yielding $\mathbf{C}_{\mathrm{BCRLB}} = \mathbf{F}_{\mathrm{total}}^{-1}$.

\subsection{ISAC Resource Allocation and Scaling Laws}
\label{subsec:resource_allocation}

The pilot fraction $\alpha = N_p/K \in (0, 1]$ parameterizes the fundamental ISAC trade-off. Under the constant-energy baseline, pilot energy scales as $E_{\mathrm{pilot}} = \alpha E_{\mathrm{total}}$, while the tracking update interval scales inversely as $\Delta t = T_{\mathrm{frame}}/\alpha$. This $\alpha$-parameterization constitutes a joint-scaling baseline that couples pilot energy fraction and tracking update rate for analytical tractability; alternative protocol designs may decouple these degrees of freedom without altering the qualitative trade-off structure. These relationships establish two distinct scaling regimes with different exponents, reflecting the underlying physics of phase estimation versus dynamic state tracking.

\textit{Residual Phase Noise Scaling ($\sigma^2_{\phi,c,\mathrm{res}} \propto \alpha^{-1}$):}
The tracking performance of the common phase noise is governed by the pilot loop SNR. Based on the Cramér-Rao lower bound for static phase estimation in AWGN~\cite{kay1993fundamentals}, the Fisher information scales linearly with pilot energy:
\begin{equation}
    \mathcal{J}_{\phi\phi} = \frac{2 E_{\mathrm{pilot}}}{N_0}.
    \label{eq:fisher_phase}
\end{equation}
Accounting for loop implementation loss $\Gamma_{\mathrm{loop}} \geq 1$, the residual phase variance is:
\begin{equation}
    \sigma^2_{\phi,c,\mathrm{res}}(\alpha) = \frac{\Gamma_{\mathrm{loop}}}{\mathcal{J}_{\phi\phi}} = \frac{\Gamma_{\mathrm{loop}} N_0}{2 E_{\mathrm{pilot}}} = \frac{\Gamma_{\mathrm{loop}} N_0}{2 \alpha E_{\mathrm{total}}} \triangleq \frac{C_{\mathrm{PN}}}{\alpha},
    \label{eq:pn_scaling}
\end{equation}
where $C_{\mathrm{PN}} = \Gamma_{\mathrm{loop}} N_0 / (2 E_{\mathrm{total}})$ aggregates the loop implementation loss and baseline SNR. The $\alpha^{-1}$ exponent arises directly from the linear relationship between Fisher information and observation energy---a fundamental result in estimation theory. This scaling dominates the high-overhead regime ($\alpha > 0.1$).

\textit{DSE Mismatch Scaling ($\sigma^2_{\phi,\mathrm{DSE}} \propto \alpha^{-5}$):}
The dynamic state-estimation error (DSE), induced by Doppler squint, manifests as a quadratic phase error trajectory corresponding to residual acceleration mismatch. To capture the orbital dynamics of LEO satellites, we model the target state evolution using a \emph{Constant Acceleration (CA)} kinematic model~\cite{bar2001estimation}:
\begin{equation}
    \mathbf{x}(t) = [r(t), \, v(t), \, a(t)]^T, \quad \dot{\mathbf{x}}(t) = \mathbf{A}\mathbf{x}(t) + \mathbf{w}(t),
    \label{eq:ca_model}
\end{equation}
where $r(t)$ is range, $v(t)$ is range rate, $a(t)$ is acceleration, and $\mathbf{w}(t) = [0, \, 0, \, w_j(t)]^T$ represents process noise driven by white jerk $w_j(t)$ with spectral density $q_j$, accounting for unmodeled high-order orbital perturbations.

Discretizing over the tracking update interval $\Delta t$, the position error variance $[\mathbf{Q}]_{1,1}$ is derived by propagating jerk noise through three integrations (jerk $\to$ acceleration $\to$ velocity $\to$ position). The impulse response from jerk to position is $h(t) = t^2/2$ for $t \geq 0$, yielding:
\begin{equation}
    \sigma^2_r(\Delta t) = q_j \int_0^{\Delta t} \left(\frac{\tau^2}{2}\right)^2 d\tau = q_j \int_0^{\Delta t} \frac{\tau^4}{4} \, d\tau = \frac{q_j \Delta t^5}{20}.
    \label{eq:pos_variance_ca}
\end{equation}
This $\Delta t^5$ dependence is characteristic of the CA model with jerk noise, distinguishing it from the $\Delta t^3$ scaling of constant-velocity models.

Mapping range error to phase error via $\tilde{\phi} = (4\pi f_c/c)\tilde{r}$, consistent with the two-way time-transfer convention in~\eqref{eq:rmse_range}, and substituting $\Delta t = T_{\mathrm{frame}}/\alpha$:
\begin{equation}
    \sigma^2_{\phi,\mathrm{DSE}}(\alpha) = \left(\frac{4\pi f_c}{c}\right)^2 \sigma^2_r = \left(\frac{4\pi f_c}{c}\right)^2 \frac{q_j}{20} \left(\frac{T_{\mathrm{frame}}}{\alpha}\right)^5 \triangleq \frac{C_{\mathrm{DSE}}}{\alpha^5},
    \label{eq:dse_scaling}
\end{equation}
where $\sigma^2_{\phi,\mathrm{DSE}}(\alpha)$ denotes the phase-mismatch variance (in $\mathrm{rad}^2$), and $C_{\mathrm{DSE}}$ encapsulates the carrier frequency, jerk noise spectral density, and frame duration. Since DSE originates from signal-model mismatch (residual quadratic phase) rather than thermal noise, the equivalent noise PSD $\sigma^2_{\mathrm{DSE}}(\alpha) \triangleq N_0 \sigma^2_{\phi,\mathrm{DSE}}(\alpha)$ is adopted as a first-order surrogate consistent with the misspecified Cramér-Rao bound (MCRB), valid in the small-mismatch regime where $\sigma^2_{\phi,\mathrm{DSE}} \ll 1$~rad$^2$. This normalization ensures that $\sigma^2_{\mathrm{DSE}}/N_0 = \sigma^2_{\phi,\mathrm{DSE}}$ remains dimensionless in~\eqref{eq:sinr_f}. The steep $\alpha^{-5}$ power law creates a performance cliff in the low-overhead regime ($\alpha \ll 0.1$), establishing an operationally infeasible region under the adopted baseline.

\textit{Physical Interpretation of Scaling Law Asymmetry:}
The dramatic difference between the exponents ($-1$ versus $-5$) has fundamental physical origins. Phase noise estimation is a \emph{static} parameter problem where Fisher information scales linearly with observation energy. DSE, by contrast, arises from \emph{dynamic} state evolution where prediction error accumulates polynomially with the prediction horizon through the CA kinematic model. These scaling relations are validated by the slopes $m = -1$ and $m = -5$ measured in log-log plots presented in Section~IV.

\textit{Regime Crossover and Design Implications:}
The transition between phase-noise-dominated and DSE-dominated regimes occurs when the two phase-error variances equalize. Setting $\sigma^2_{\phi,c,\mathrm{res}}(\alpha^*) = \sigma^2_{\phi,\mathrm{DSE}}(\alpha^*)$:
\begin{equation}
    \frac{C_{\mathrm{PN}}}{\alpha^*} = \frac{C_{\mathrm{DSE}}}{(\alpha^*)^5} \quad \Rightarrow \quad \alpha^* = \left(\frac{C_{\mathrm{DSE}}}{C_{\mathrm{PN}}}\right)^{1/4}.
    \label{eq:alpha_crossover}
\end{equation}
The $1/4$ exponent emerges from $1/(5-1)$. This predictor yields $\alpha^* \approx 0.16$ for typical LEO-ISL parameters. The crossover $\alpha^*$ decreases with improved oscillator quality (smaller $C_{\mathrm{PN}}$) and increases with faster dynamics (larger $C_{\mathrm{DSE}}$).

\textit{Alpha-Parameterized Performance:}
Substituting the scaling laws, the spectral efficiency surrogate becomes:
\begin{equation}
    C_{\mathrm{J}}(\alpha) = \log_2\left(1 + \frac{\mathrm{SNR}_0 G_{\mathrm{sig,avg}} e^{-C_{\mathrm{PN}}/\alpha}}{1 + \mathrm{SNR}_0 G_{\mathrm{sig,avg}} \Gamma_{\mathrm{eff,total}} + C_{\mathrm{DSE}}/\alpha^5}\right),
    \label{eq:capacity_alpha}
\end{equation}
where the numerator captures phase-noise-induced coherence loss, and $C_{\mathrm{DSE}}/\alpha^5 = \sigma^2_{\phi,\mathrm{DSE}}(\alpha)$ is the dimensionless DSE term per~\eqref{eq:dse_scaling}. The same $C_{\mathrm{PN}}, C_{\mathrm{DSE}}$ apply to the sensing BCRLB.

The net communication rate accounting for pilot overhead is:
\begin{equation}
    R_{\mathrm{net}}(\alpha) = (1 - \alpha) \cdot C_{\mathrm{J}}(\alpha).
    \label{eq:net_rate}
\end{equation}

\textit{Pareto Optimization:}
The bi-objective ISAC system seeks to maximize throughput while minimizing sensing RMSE:
\begin{equation}
    \max_{\alpha} \; R_{\mathrm{net}}(\alpha), \quad \min_{\alpha} \; \mathrm{RMSE}(\alpha) = \sqrt{[\mathbf{C}_{\mathrm{BCRLB}}(\alpha)]_{1,1}}, \quad \text{s.t.} \; \alpha \in (0, 1].
    \label{eq:pareto_problem}
\end{equation}
The function $R_{\mathrm{net}}(\alpha)$ is non-convex due to the product of linear $(1-\alpha)$ and exponential $e^{-C_{\mathrm{PN}}/\alpha}$ terms, precluding closed-form solution; the Pareto frontier is generated via parameter scan.

\textit{Boundary Behavior:}
At $\alpha \to 0$, both $\sigma^2_{\phi,c,\mathrm{res}} \to \infty$ and $\sigma^2_{\phi,\mathrm{DSE}} \to \infty$ cause performance divergence: $\mathrm{RMSE} \to \infty$ and $R_{\mathrm{net}} \to 0$. This defines an operationally infeasible region under the adopted baseline, requiring minimum overhead $\alpha_{\min}$ to bootstrap the system.

\section{Numerical Results}
\label{sec:numerical_results}

This section validates the theoretical framework developed in Sections~\ref{sec:system_model}--\ref{sec:performance_limits} through Monte Carlo simulations. Key results include MIMO scaling laws, capacity saturation, and the $\alpha^{-1}$/$\alpha^{-5}$ scaling law crossover.

\subsection{Simulation Setup}
\label{subsec:simulation_setup}

Table~\ref{tab:sim_params} summarizes simulation parameters for two hardware configurations. The \textit{Low-Cost} tier represents near-term payloads where PA efficiency is prioritized over linearity, yielding $C_{\mathrm{sat}} \approx 2.8$~bits/s/Hz. The \textit{Baseline} tier assumes wideband DPD linearization, reducing $\Gamma_{\mathrm{PA}}$ to $-22$~dB and achieving $C_{\mathrm{sat}} \approx 7.4$~bits/s/Hz; however, implementing DPD over $B = 20$~GHz requires gigasample-rate feedback and high-speed DSP that remain challenging for SWaP-constrained LEO payloads. The Baseline should therefore be interpreted as an optimistic upper bound, while the Low-Cost tier is closer to current THz ISL hardware. Unless otherwise noted, numerical results employ the Baseline configuration.

\begin{table}[t]
\centering
\caption{Simulation Parameters}
\label{tab:sim_params}
\begin{tabular}{lccc}
\hline
\textbf{Parameter} & \textbf{Symbol} & \textbf{Low-Cost} & \textbf{Baseline} \\
\hline
Carrier frequency & $f_c$ & \multicolumn{2}{c}{140 GHz} \\
Bandwidth & $B$ & \multicolumn{2}{c}{20 GHz} \\
Array elements (Tx/Rx) & $N_t = N_r$ & \multicolumn{2}{c}{64} \\
Element spacing & $d$ & \multicolumn{2}{c}{$\lambda_c/2$} \\
Steering angle & $\theta_0$ & \multicolumn{2}{c}{$30^\circ$} \\
Aperture-wavelength ratio & $L_{\mathrm{ap}}/\lambda_c$ & \multicolumn{2}{c}{32} \\
\hline
\multicolumn{4}{c}{\textit{Hardware Impairment Parameters}} \\
\hline
PA distortion (EVM) & $\Gamma_{\mathrm{PA}}$ & $-8$ dB & $-22$ dB$^\dagger$ \\
ADC resolution & $b_{\mathrm{ADC}}$ & \multicolumn{2}{c}{7 bits} \\
I/Q imbalance (IRR) & $\Gamma_{\mathrm{IQ}}$ & \multicolumn{2}{c}{$-20$ dB} \\
Phase shifter bits & $b_\phi$ & \multicolumn{2}{c}{4 bits} \\
RMS timing jitter & $\sigma_t$ & \multicolumn{2}{c}{50 fs} \\
Loop loss & $\Gamma_{\mathrm{loop}}$ & \multicolumn{2}{c}{3 dB} \\
\hline
\multicolumn{4}{c}{\textit{Derived Quantities}} \\
\hline
Beam squint factor & $\kappa$ & \multicolumn{2}{c}{1.14} \\
Total HW distortion & $\Gamma_{\mathrm{eff,total}}$ & $-7.7$ dB & $-22$ dB \\
Saturation capacity & $C_{\mathrm{sat}}$ & $\approx 2.8$ & $\approx 7.4$ bits/s/Hz \\
\hline
\multicolumn{4}{l}{\footnotesize $^\dagger$Assumes digital predistortion (DPD) linearization.}
\end{tabular}
\end{table}

The frequency-domain Whittle-FIM~\eqref{eq:whittle_fim} provides computational efficiency over exact time-domain Cholesky decomposition. Numerical verification across 100 parameter configurations ($L_{\mathrm{ap}}/\lambda_c \in [3, 25]$, $B/f_c \in [0.02, 0.15]$) confirms that the approximation error remains below $2\%$, justifying its use for BCRLB computation in the considered wideband regime.

\begin{figure}[t]
    \centering
    \includegraphics[width=0.9\columnwidth]{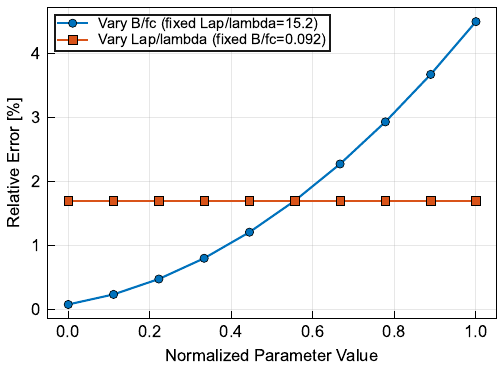}
    \caption{Whittle-FIM approximation validation: sensitivity analysis showing error increases with fractional bandwidth $B/f_c$ but remains insensitive to normalized aperture.}
    \label{fig:threshold_slices}
\end{figure}

The sensitivity analysis in Fig.~\ref{fig:threshold_slices} reveals that approximation accuracy degrades with increasing fractional bandwidth $B/f_c$ but remains insensitive to normalized aperture $L_{\mathrm{ap}}/\lambda_c$. This behavior is consistent with the Whittle approximation's derivation, which assumes narrowband signals ($B \ll f_c$) but imposes no constraint on aperture size. For the baseline configuration ($B/f_c = 0.143$), the relative error is approximately $1.8\%$, validating the use of~\eqref{eq:whittle_fim} for BCRLB computation.

\textit{RSM Comb Spectrum:} The RSM-induced colored noise exhibits a comb-like spectrum with narrow peaks at symbol-rate harmonics. Whittle's approximation can be sensitive to such spectral fine structure if the signal spectrum is highly non-uniform. In the THz ISL setting, the probing waveform is designed to be approximately flat over the occupied band, and the RSM spectral lines are narrow compared with the integration bandwidth, so the effective FIM averages over many peaks and valleys. This explains the $<2\%$ discrepancy observed in Fig.~\ref{fig:threshold_slices}. Extending this analysis to strongly structured waveforms remains an open direction.

\subsection{MIMO Scaling Laws}
\label{subsec:mimo_scaling}

\begin{figure}[t]
    \centering
    \includegraphics[width=0.9\columnwidth]{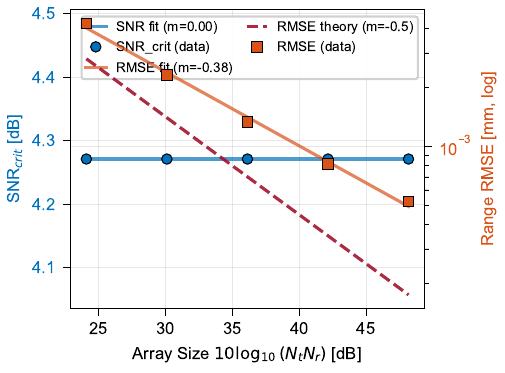}
    \caption{MIMO scaling laws: critical SNR exhibits scale invariance (slope $m \approx 0$), while sensing RMSE decreases with slope $m = -0.42$.}
    \label{fig:mimo_scaling}
\end{figure}

Fig.~\ref{fig:mimo_scaling} illustrates communication and sensing performance scaling with array size $N$ (where $N_t = N_r = N$).

\textit{Sensing RMSE:} The ranging RMSE decreases with fitted slope $m = -0.42$, slightly below the theoretical $m = -0.50$ from~\eqref{eq:awgn_baseline}. This deviation arises from beam squint: larger apertures increase $\kappa$, reducing effective gain at band edges.

\textit{Critical SNR:} The critical SNR exhibits \textbf{scale invariance} (slope $m \approx 0$), arising from the ``directional distortion'' phenomenon analyzed in Section~\ref{subsec:comm_performance}: under CE transmission, PA distortion beamforms coherently with the signal, causing both to scale as $N_t N_r$. The resulting constant signal-to-distortion ratio stabilizes the distortion-limited transition point regardless of array size.
\subsection{Communication Performance}
\label{subsec:comm_performance}

\begin{figure}[t]
    \centering
    \includegraphics[width=0.9\columnwidth]{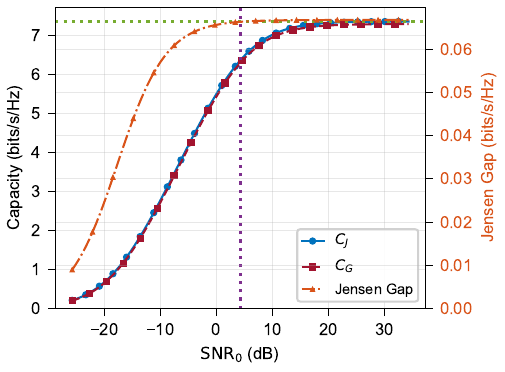}
    \caption{Capacity versus SNR with saturation ceiling $C_{\mathrm{sat}} \approx 7.4$~bits/s/Hz.}
    \label{fig:capacity_vs_snr}
\end{figure}

Fig.~\ref{fig:capacity_vs_snr} presents capacity versus $\mathrm{SNR}_0$, comparing the Jensen bound $C_{\mathrm{J}}$, exact Gaussian capacity $C_G$, and the hardware-imposed ceiling.

\textit{Saturation Ceiling:} The capacity saturates at $C_{\mathrm{sat}} \approx 7.4$~bits/s/Hz for the Baseline configuration, matching~\eqref{eq:saturation}:
\begin{equation}
    C_{\mathrm{sat}} = \log_2\left(1 + \frac{e^{-\sigma^2_{\phi,c,\mathrm{res}}}}{\Gamma_{\mathrm{eff,total}}}\right) = \log_2\left(1 + \frac{0.99}{0.006}\right) \approx 7.4~\text{bits/s/Hz}.
    \notag
\end{equation}
Here $\Gamma_{\mathrm{eff,total}} = 0.006$ ($-22$~dB) corresponds to the Baseline tier with DPD. For the Low-Cost tier ($\Gamma_{\mathrm{eff,total}} = 0.17$, i.e., $-7.7$~dB), the ceiling drops to $C_{\mathrm{sat}} = \log_2(1 + 0.99/0.17) \approx 2.8$~bits/s/Hz. With $B = 20$~GHz, the achievable rates are $148$~Gbps (Baseline) or $56$~Gbps (Low-Cost). While the Baseline curve illustrates achievable performance with aggressive DPD, the Low-Cost curve is arguably closer to what is feasible in current THz ISL hardware given SWaP constraints.

\textit{Jensen Gap:} The gap between $C_{\mathrm{J}}$ and $C_G$ peaks at $0.065$~bits/s/Hz near the critical SNR transition, representing $<1\%$ relative error. For the Baseline wideband setting ($\kappa=1.14$ in Table~\ref{tab:sim_params}), the Jensen gap peaks at $0.065$~bits/s/Hz ($<1\%$), indicating that~\eqref{eq:jensen_bound} is a tight surrogate in the considered operating regime; the Taylor-based discussion in~\eqref{eq:jensen_gap} is used only for qualitative scaling insight.

The PA nonlinearity contributes $88.7\%$ of the total distortion budget $\Gamma_{\mathrm{eff,total}}$, confirming that the system is fundamentally PA-efficiency-limited. Improving ADC resolution beyond 7 bits yields negligible benefit, while even modest PA linearization (e.g., digital predistortion) could substantially raise $C_{\mathrm{sat}}$.

\textit{Remark (Directional Distortion):} The PA-dominated distortion budget has an important implication for MIMO scaling. Under constant-envelope (CE) transmission, each per-element signal has the form $x_n = e^{j\phi_n}s$, where $|x_n|$ is constant. A memoryless PA nonlinearity produces distortion $d_n = g(|x_n|)x_n \propto x_n$, implying that the distortion vector $\mathbf{d}$ is proportional to the beamforming vector $\mathbf{a}_t$. Consequently, the distortion beamforms \textit{coherently} with the signal toward the receiver:
\begin{equation}
    \mathbf{a}_t^H \mathbf{d} \propto \mathbf{a}_t^H \mathbf{a}_t \cdot s = N_t \cdot s.
    \notag
\end{equation}
Both signal and distortion power scale as $N^2$, yielding a constant Signal-to-Distortion Ratio (SDR) independent of array size. This ``directional distortion'' phenomenon explains why $C_{\mathrm{sat}}$ represents a fundamental ceiling that cannot be overcome by simply adding more antennas---in contrast to uncorrelated impairments (e.g., thermal noise) which benefit from massive MIMO averaging. The proportionality $d_n \propto x_n$ assumes identical memoryless PA characteristics across antennas (i.e., the same nonlinear mapping $g(\cdot)$ under CE drive). With per-PA parameter mismatch, the distortion becomes only partially aligned with $\mathbf{a}_t$, and the post-combining distortion power can be decomposed into a coherent and an incoherent part. Hence, the scale-invariant SDR and the resulting saturation ceiling represent a worst-case (fully correlated) impairment model; partial decorrelation relaxes but does not eliminate distortion-induced saturation in practical array sizes.

\subsection{Sensing Performance and ISAC Trade-offs}
\label{subsec:sensing_isac}

\begin{figure}[t]
    \centering
    \includegraphics[width=0.9\columnwidth]{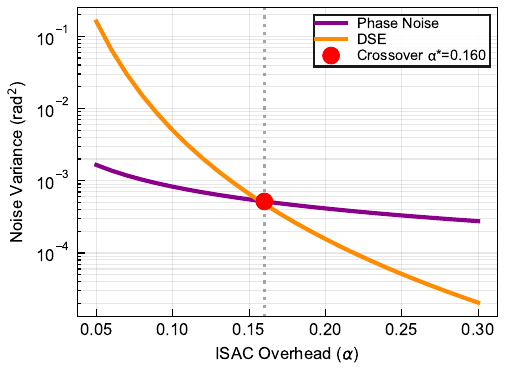}
    \caption{Scaling law verification: PN ($\alpha^{-1}$), DSE ($\alpha^{-5}$), crossover at $\alpha^*=0.16$.}
    \label{fig:pn_dse_crossover}
\end{figure}

Fig.~\ref{fig:pn_dse_crossover} provides direct verification of the $\alpha$-dependent scaling laws derived in Section~\ref{subsec:resource_allocation}.

In log-log scale, the phase noise variance exhibits slope $m = -1$ (confirming~\eqref{eq:pn_scaling}) while DSE variance exhibits slope $m = -5$ (confirming~\eqref{eq:dse_scaling}). The crossover occurs at $\alpha^* = 0.160$, matching the theoretical prediction from~\eqref{eq:alpha_crossover}:
\begin{equation}
    \alpha^* = \left(\frac{C_{\mathrm{DSE}}}{C_{\mathrm{PN}}}\right)^{1/4} \approx 0.16.
    \notag
\end{equation}
This validates both the Constant Acceleration kinematic model assumption and the jerk-noise-driven $\Delta t^5$ position variance derivation~\eqref{eq:pos_variance_ca}.

\begin{figure}[t]
    \centering
    \includegraphics[width=0.9\columnwidth]{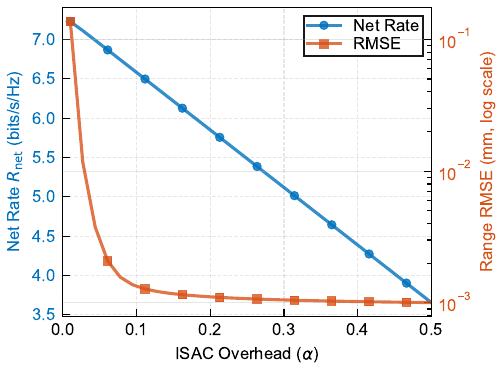}
    \caption{Net rate and RMSE versus pilot overhead $\alpha$.}
    \label{fig:alpha_combined}
\end{figure}

Fig.~\ref{fig:alpha_combined} presents the joint evolution of net rate $R_{\mathrm{net}}$ and sensing RMSE as functions of pilot overhead $\alpha$. The net rate $R_{\mathrm{net}} = (1-\alpha)C_{\mathrm{J}}(\alpha)$ decreases monotonically with $\alpha$ due to the overhead penalty. The RMSE exhibits a characteristic ``knee'' near $\alpha^* \approx 0.16$: below this threshold, DSE dominates and RMSE increases rapidly; above it, RMSE improves gradually toward the hardware floor. This confirms the operationally infeasible region under the adopted baseline---operating below $\alpha^*$ sacrifices both communication throughput and sensing accuracy.

\begin{figure}[t]
    \centering
    \includegraphics[width=0.9\columnwidth]{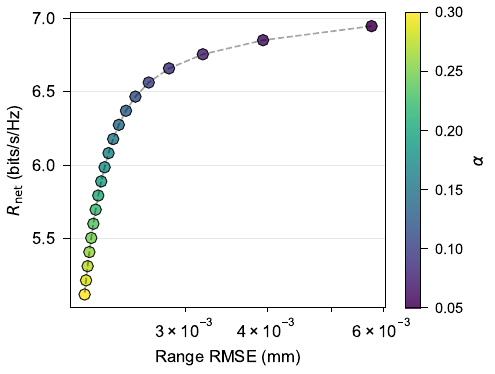}
    \caption{ISAC Pareto frontier in $(R_{\mathrm{net}}, \mathrm{RMSE})$ space.}
    \label{fig:pareto_frontier}
\end{figure}

The Pareto frontier in Fig.~\ref{fig:pareto_frontier} visualizes this trade-off in $(R_{\mathrm{net}}, \mathrm{RMSE})$ space. Key operating points along the frontier:
\begin{itemize}
    \item $\alpha = 0.05$: $R_{\mathrm{net}} \approx 6.9$~bits/s/Hz, RMSE $\approx 5.8$~$\mu$m (DSE-limited, inefficient)
    \item $\alpha = 0.10$: $R_{\mathrm{net}} \approx 6.6$~bits/s/Hz, RMSE $\approx 2.6$~$\mu$m (near-optimal)
    \item $\alpha = 0.30$: $R_{\mathrm{net}} \approx 5.1$~bits/s/Hz, RMSE $\approx 2.1$~$\mu$m (sensing-prioritized)
\end{itemize}
The RMSE of $\approx 2.6$~$\mu$m at $\alpha = 0.10$ corresponds to sub-wavelength ranging precision ($\lambda_c = 2.14$~mm at 140~GHz), enabling high-precision relative navigation for formation flying satellites.

\begin{figure*}[t]
    \centering
    \begin{subfigure}[b]{0.45\textwidth}
        \centering
        \includegraphics[width=\textwidth]{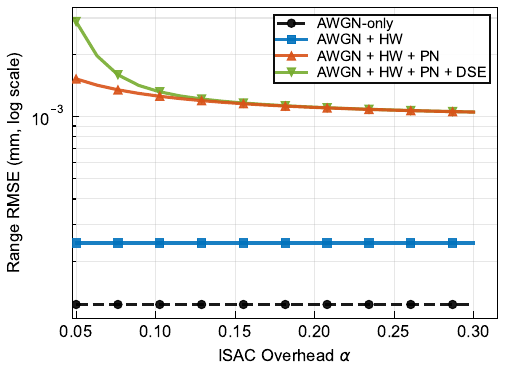}
        \caption{Cumulative impairment impact on RMSE}
        \label{fig:ablation}
    \end{subfigure}
    \hfill
    \begin{subfigure}[b]{0.45\textwidth}
        \centering
        \includegraphics[width=\textwidth]{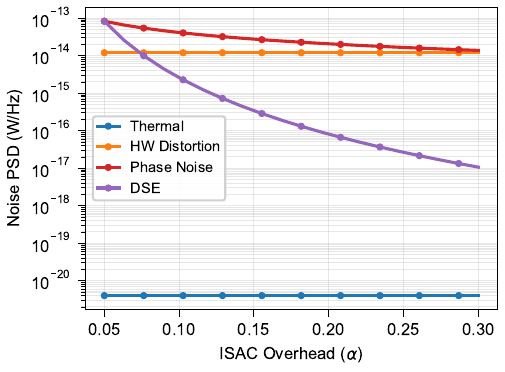}
        \caption{Noise PSD composition versus $\alpha$}
        \label{fig:noise_comp}
    \end{subfigure}
    \caption{Ablation study of impairment contributions: (a) cumulative impact of hardware distortion (HW), phase noise (PN), and dynamic state estimation error (DSE) on ranging RMSE; (b) noise power spectral density decomposition showing DSE-PN crossover at $\alpha \approx 0.16$.}
    \label{fig:ablation_combined}
\end{figure*}

Fig.~\ref{fig:ablation} isolates the impact of individual impairment sources on sensing RMSE through cumulative addition: static hardware distortions (HW) create a constant floor $\approx 3\times$ above the thermal limit; residual phase noise (PN) introduces $\alpha$-dependence with slope $m \approx -0.5$; and DSE creates the steep performance cliff at low $\alpha$. The infeasible region is thus a \emph{dynamic} phenomenon arising from CA model prediction error accumulation, not a static hardware limitation.

The spectral decomposition in Fig.~\ref{fig:noise_comp} shows DSE noise power varying over six orders of magnitude, with the crossover at $\alpha \approx 0.16$ matching the theoretical $\alpha^*$ prediction. The system transitions from DSE-dominated to PN/HW-dominated operation as $\alpha$ increases past $\alpha^*$.


\subsection{Pareto Frontier Sensitivity}
\label{subsec:pareto_sensitivity}

To provide design guidance across different system configurations, we examine how the Pareto frontier shifts with array size, hardware quality, and signal bandwidth.

\begin{figure*}[t]
    \centering
    \begin{subfigure}[b]{0.32\textwidth}
        \centering
        \includegraphics[width=\textwidth]{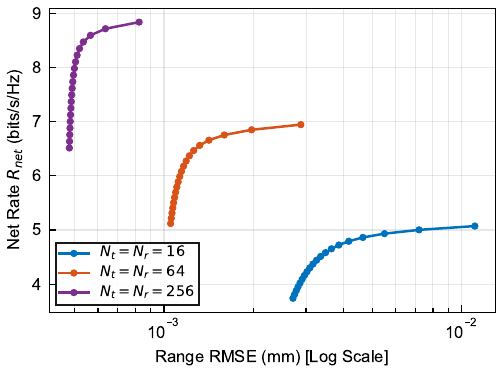}
        \caption{Array size scaling}
        \label{fig:pareto_array}
    \end{subfigure}
    \hfill
    \begin{subfigure}[b]{0.32\textwidth}
        \centering
        \includegraphics[width=\textwidth]{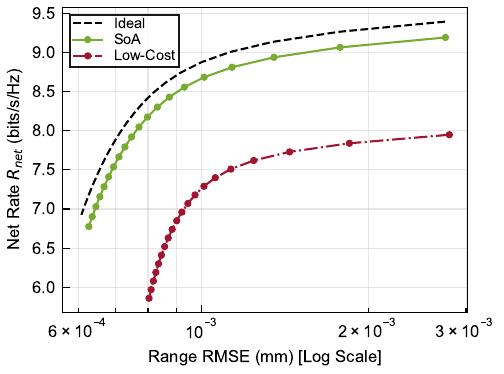}
        \caption{Hardware quality}
        \label{fig:pareto_hardware}
    \end{subfigure}
    \hfill
    \begin{subfigure}[b]{0.32\textwidth}
        \centering
        \includegraphics[width=\textwidth]{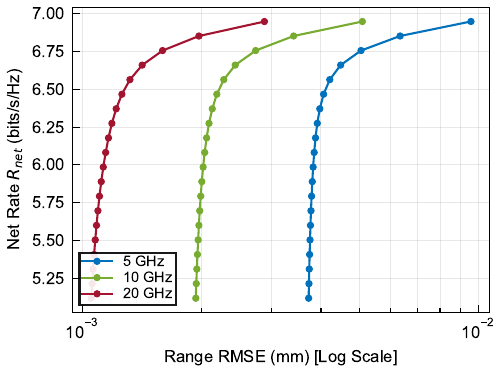}
        \caption{Signal bandwidth}
        \label{fig:pareto_bandwidth}
    \end{subfigure}
    \caption{Pareto frontier sensitivity to system parameters: (a) array size $N \in \{16, 64, 256\}$, (b) hardware quality tiers, and (c) bandwidth $B \in \{5, 10, 20\}$~GHz.}
    \label{fig:pareto_sensitivity}
\end{figure*}

\textit{Array Size Scaling:}
Fig.~\ref{fig:pareto_array} shows that increasing array size shifts the Pareto frontier toward lower RMSE and higher $R_{\mathrm{net}}$. From $N = 16$ to $N = 256$, the minimum RMSE decreases by $\approx 10\times$ (below the theoretical $16\times$ due to beam squint at larger apertures).

\textit{Hardware Quality Sensitivity:}
Fig.~\ref{fig:pareto_hardware} compares Ideal, SoA ($\Gamma_{\mathrm{eff}} = -22$~dB), and Low-Cost ($\Gamma_{\mathrm{eff}} = -8$~dB) tiers. Hardware linearization primarily benefits communication ($16\%$ capacity improvement) while providing limited sensing improvement ($<5\%$), since sensing in the operational $\alpha$ range is dominated by phase noise and DSE.

\textit{Bandwidth Trade-off:} Fig.~\ref{fig:pareto_bandwidth} shows that doubling bandwidth reduces minimum RMSE by $\approx 2\times$, consistent with $\mathrm{RMSE} \propto 1/B$. The reduced pilot overhead required to achieve sub-millimeter RMSE creates an \textbf{overhead dividend} that outweighs beam squint losses, making bandwidth expansion effective for both sensing and communication.

\subsection{Parameter Sensitivity and Design Guidelines}
\label{subsec:sensitivity}
To translate the theoretical findings into actionable engineering guidelines, we examine how key hardware and environmental parameters affect the achievable performance.

\begin{figure}[t]
    \centering
    \includegraphics[width=0.9\columnwidth]{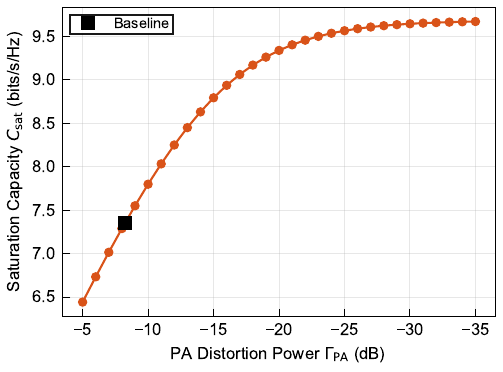}
    \caption{Saturation capacity $C_{\mathrm{sat}}$ versus PA distortion level $\Gamma_{\mathrm{PA}}$, showing diminishing returns once $\Gamma_{\mathrm{PA}} \lesssim -25$~dB.}
    \label{fig:sensitivity_pa}
\end{figure}

\textit{Capacity sensitivity to PA linearity:}
Fig.~\ref{fig:sensitivity_pa} quantifies the impact of PA nonlinearity on the saturation capacity $C_{\mathrm{sat}}$. The Low-Cost tier ($\Gamma_{\mathrm{PA}} = -8$~dB, no DPD) achieves only $C_{\mathrm{sat}} \approx 2.8$~bits/s/Hz, while the Baseline tier ($\Gamma_{\mathrm{PA}} = -22$~dB, with wideband DPD) reaches $\approx 7.4$~bits/s/Hz, corresponding to a $\approx 2.6\times$ improvement in spectral efficiency. Beyond $\Gamma_{\mathrm{PA}} \approx -25$~dB, the capacity gain rapidly diminishes as secondary impairments (ADC quantization, IQ imbalance, LO phase noise) dominate the aggregate distortion $\Gamma_{\mathrm{eff,total}}$. This identifies an ``investment sweet spot'' for PA linearity around $-20$ to $-25$~dB: pushing DPD significantly beyond this range yields marginal capacity returns while incurring substantial SWaP cost.

\textit{Overhead sensitivity to orbital dynamics (no extra figure):}
The optimal pilot overhead $\alpha^*$ is governed by the PN--DSE crossover condition in~\eqref{eq:alpha_crossover}. Since the DSE variance scales as $\sigma_{\mathrm{DSE}}^2 \propto q_j \alpha^{-5}$ while the PN variance scales as $\sigma_{\phi,c,\mathrm{res}}^2 \propto \alpha^{-1}$, we obtain the design rule $\alpha^* \propto q_j^{1/4}$. In the baseline scenario, this yields $\alpha^* \approx 0.16$; reducing the jerk noise intensity by two orders of magnitude (e.g., via improved orbit prediction or cooperative ranging) lowers the required overhead to $\alpha^* < 0.08$, effectively more than doubling the fraction of resources available for data transmission.

\section{Conclusion}
This work has established theoretical performance limits of THz-ISL MIMO-ISAC systems operating under per-element constant-envelope (CE) constraints---a design necessity for power-limited satellite platforms. A unified dual-scale framework integrates communication-side coherence loss (Bussgang model) with sensing-side residual noise (Whittle-FIM), linked through a spectral consistency principle that ensures phase noise is counted exactly once.

Three principal findings emerge. First, communication capacity is bounded by a hardware-determined ceiling ($C_{\mathrm{sat}} \approx 7.4$~bits/s/Hz with DPD, or $\approx 2.8$~bits/s/Hz without), confirming that transmit power becomes a diminishing resource in distortion-limited channels. Second, MIMO scaling laws diverge: while sensing RMSE improves as $1/\sqrt{N_t N_r}$, the critical SNR exhibits scale invariance, indicating that the distortion-limited transition point remains stable regardless of the array size.
Third, the steep $\alpha^{-5}$ scaling of DSE variance---a constraint-driven consequence of jerk-noise accumulation---creates an operationally infeasible region at $\alpha < \alpha^* \approx 0.16$ under the adopted baseline. The closed-form crossover predictor $\alpha^* = (C_{\mathrm{DSE}}/C_{\mathrm{PN}})^{1/4}$ provides design guidance for LEO constellations.

These constraint-driven findings indicate that future THz-ISL designs should prioritize hardware linearity and phase synchronization over raw transmit power or aperture size. The derived Pareto frontier provides a quantitative framework for resource allocation in next-generation satellite constellations.

\bibliographystyle{IEEEtran}
\bibliography{references}

\vspace{-1.2cm} 

\begin{IEEEbiography}
[{\includegraphics[width=1in,height=1.25in,clip]{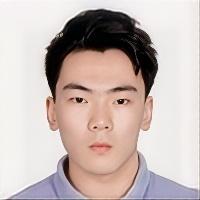}}]{Haofan Dong(hd489@cam.ac.uk)}
 is a Ph.D. student in the Internet of Everything (IoE) Group, Department of Engineering, University of Cambridge, UK. He received his MRes from CEPS CDT based in UCL in 2023. His research interests include integrated sensing and communication (ISAC), space communications, and THz communications.
\end{IEEEbiography}
\vspace{-1.2cm} 

\begin{IEEEbiography}[{\includegraphics[width=1in,height=1.25in,clip]{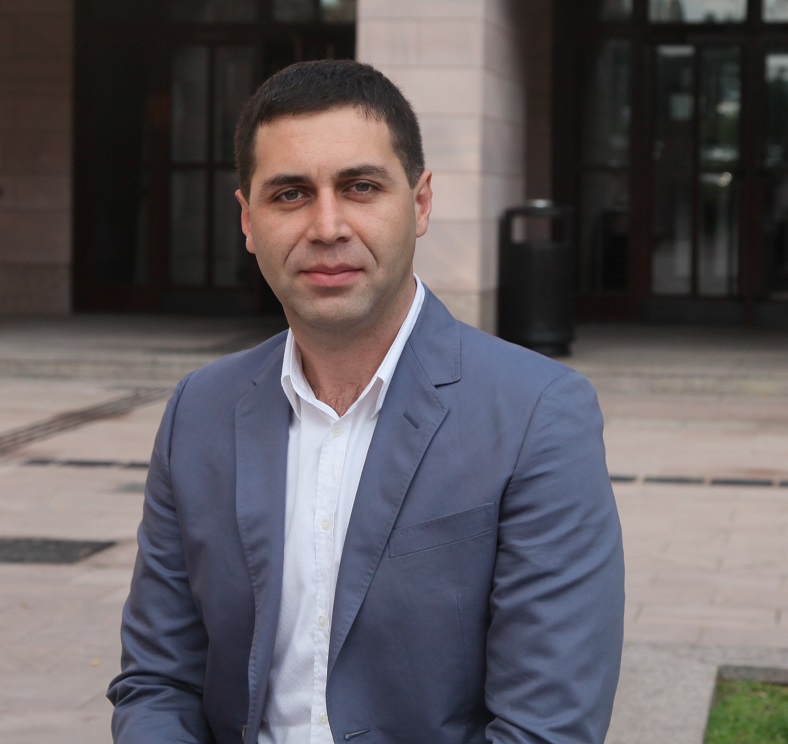}}]{Ozgur B. Akan(oba21@cam.ac.uk) }
received his Ph.D. degree from the School of Electrical and Computer Engineering, Georgia Institute of Technology, Atlanta, in 2004. He is currently the Head of the Internet of Everything (IoE) Group, Department of Engineering, University of Cambridge, and the Director of the Centre for NeXt-Generation Communications (CXC), Koç University. His research interests include wireless, nano-, and molecular communications, and the Internet of Everything.
\end{IEEEbiography}
\vspace{-1.2cm} 

\end{document}